\documentclass{cs19proc}

\editors{G.~A. Feiden}
\publisher{Zenodo}
\conference{The 19th Cambridge Workshop on Cool Stars, Stellar Systems, and the Sun}
\conferencedate{2016}

\title{An analytical test case for dust dynamics during a shock-wave passage}
\author{Lars Mattsson}

\affiliation{Nordita, Royal Institute of Technology and Stockholm University, Roslagstullsbacken 23, SE-106 91, Stockholm, Sweden}

\shorttitle{An analytical test case for dust dynamics during a shock-wave passage}
\shortauthors{Lars Mattsson}

\abs{An exact solution of a forced Burgers' equation representing the dynamics of a "dust fluid" in a one-dimensional flow is presented. The test case considered starts with a steady (time independent) two-fluid flow in one dimension, where the two fluid components represent gas and dust. It is then assumed that a shock wave travels through the gas at a constant speed and without radiative energy losses and diffusion. Then, adopting a constant stopping time for the dust particles in the dust fluid (mono-dispersed grain sizes), the equation of motion for the dust fluid can be transformed into a simple ordinary differential equation, which is satisfied by the Wright omega function. Implications for the formation of detached shells around carbon stars are briefly discussed.}

\begin{document}

\maketitle

\section{Introduction}
Particles in a fluid flow will experience a drag force, which depends on the properties of the flow as well as the particles. If the fluid is a gas, and the mean-free path $\ell_{\rm MFP}$ of the gas particles is long, this is known as Epstein drag \citep{Epstein24}, in which the force is inversely proportional to the radius of the particle. At the other end, where $\ell_{\rm MFP}$ is short, small particles experiences drag in the so-called Stokes regime, which yields a drag-force law where the force is inversely proportional to the surface area of a spherical particle particle. In most astrophysical contexts, e.g., dust particles in the gas of the interstellar medium (ISM), the relevant regime is the Epstein regime. The present work is therefore limited to this regime.

Due to the fact that there is never perfect velocity coupling between gas and dust, the response of these two phases to a shock wave should be different. The dust is expected to lag behind the gas and the shock profile measured in the dust should be different from that in the gas phase. Formation of detached shells around carbon rich AGB stars (C stars) is one example of where this effect may occur. The current understanding of the detached-shell phenomenon is that  such structures form as a result of a sudden increase in the wind velocity following a thermal-pulse event (TPE). The subsequent wind-wind interaction leads to a nearly stationary step-like shock propagating outwards with an almost constant velocity, which is less than that of the fast (post TPE) wind, but greater than that of the slow (pre TPE) wind \citep{Mattsson07,Steffen00}. Models of detached-shell formation which assume velocity coupling between the gas and dust phases \citep[e.g.,][]{Mattsson07} imply that the density profile, as well as the shock profile, has the same shape for both gas and dust. This cannot be entirely correct, since the response of the ``dust fluid'' to the shock is expected to be slower than the response of the gas.

The aim of this work is to finding a solution to an idealised travelling-wave/shock problem for a gas/dust fluid in Eulerian coordinates. An exact expression for the velocity profile of the dust provides a test case for implementation of Epstein drag in numeric hydrodynamic simulations.

\section{The two-fluid model of dust in a gas flow}
\subsection{Equation of motion}
The dust phase can be assumed to behave as an inviscid, pressure-less fluid and couple the gas flow via a drag-force given by an Epstein law. The EOM (in non-conservative form) is then the inviscid Burgers' equation with a forcing term,
\begin{equation}
f(x,t) = {u-u_{\rm d}\over t_{\rm s}},
\end{equation}
where $t_{\rm s}$ is the so-called stopping time, i.e., the time it takes before a dust grain has accelerated (or decelerated) to same velocity as the gas flow (in case of a steady laminar flow). The stopping time depends on the size and density of the grain as well as the gas density,
\begin{equation}
t_{\rm s} = {\rho_{\rm gr}\over\rho}{a\over \langle v_{\rm th}\rangle}
\end{equation}
where $a$ is the grain radius (assuming spherical grains), $\rho_{\rm gr}$ is the bulk material density of the grain and $\langle v_{\rm th}\rangle$ is the mean thermal speed of the gas particles. Under many conditions it is legitimate to assume $t_{\rm s}$ is constant for a given grain size. In such case, we seek the solution to the following initial-value problem, 
\begin{equation}
{du_{\rm d}\over dt}={\partial u_{\rm d}\over \partial t} + u_{\rm d}{\partial u_{\rm d}\over \partial x} = {u-u_{\rm d}\over t_{\rm s}},\quad x\in \mathbb{R},\quad t>0,
\end{equation}
\begin{displaymath}
u_{\rm d}(a,x,0)= u(a,x,0) = u_{0}(a), \quad x\in \mathbb{R},\quad a>0,
\end{displaymath}
where $u_{\rm d} = u_{\rm d}(a,x,t)$ and $t_{\rm s} = t_{\rm s}(a)$. The statement of the problem is the same for spherical symmetry since the ``geometrical terms" are removed when the continuity equation is subtracted from the EOM on conservative form (momentum conservation). 

\subsection{Solutions for a simple shockwave}
\label{solutions}
Consider a one-dimensional propagating shockwave, which corresponds to a simple step function in $u$, moving from left to right with a velocity $s$. Then, assume $u_{\rm d}(a,x,t) = u_{\rm d}(a,x-s\,t)$ (travelling-wave solution) and $s > u_2 > u_1=$~constant, where $u_2$ is the velocity of the gas after the shock and $u_1$ is the pre-shock velocity. Before the passage of the shock $u_{\rm d} = u_1$, i.e., the dust is assumed to have coupled to flow. Introducing the dimensionless variables,
\begin{equation}
\varpi = {u-u_{\rm d}\over s-u} \ge 0,\quad \eta = {x-s\,t\over t_{\rm s}\,(s-u)},
\end{equation}
one can easily show that the EOM for the post-shock flow for a given $a$ reduces to
\begin{equation}
\label{wrighteq}
{d\varpi\over d\eta} = {\varpi\over \varpi +1},
\end{equation}
which has an exact, real-valued solution known as the Wright omega function. This function is related to the Lambert $W$-function via an exponential variable transformation \citep{Corless96,Corless02}, 
\begin{equation}
\omega(z) = W_{\left\lceil{\Im(z)-\pi\over 2\pi}\right\rceil}(e^z).
\end{equation}
The Lambert $W$ function is multifunction with countably infinite number of branches, but for $z\in \mathbb{R}$ only the 0 and $-1$ branches of $W_k$ are relevant, as in the formula above. The general solution to eq. (\ref{wrighteq}) is
\begin{equation}
\varpi(\eta) = \omega(\eta+C) = W_0(e^{\eta+C})
\end{equation}
where $C$ is a constant of integration. Back-transformation to physical variables yields
\begin{equation}
\label{physsol}
u_{\rm d} = \omega\left[ {x-s\,t + x_0\over t_{\rm s} (s-u_2)}\right](u_2-s) + u_2,
\end{equation}
where $x_0$ is a translation parameter to be chosen such that $u_{\rm d}$ before the shock connects with the velocity profile after the shock, i.e., the solution above. 

The case $u_2=s$ is somewhat special. The transformed equation above would have a singularity at $u_2 = s$, so this case must be handled separately. Assume $u_2 = s$ and let $\Delta u = u_2-u_{\rm d}$, $\xi = x-s\,t$. Then, 
\begin{equation}
\Delta u\,{d\Delta u\over d\xi} = {\Delta u\over t_{\rm s}},
\end{equation}
which yields two solutions,
\begin{equation}
\Delta u = {\xi\over t_{\rm s}} + C \quad {\rm and} \quad \Delta u = 0,
\end{equation}
where $C$ is a constant of integration which is normally equal to zero in the present context. The trivial solution $\Delta u = 0$ is irrelevant (no shock). In the limit $u_2\to s$, $u_{\rm d}$ will approach an interesting case corresponding to an exactly linear acceleration of the dust until $u_{\rm d} = u_2$ (see Fig. \ref{ud_s}). This seemingly unphysical solution is the result of the dust being repeatedly kicked by the shock until it reaches $u_{\rm d} = u_2 = s$. When the dust is kicked forward it is immediately kicked again as the shock catches up. The momentum gained from each kick is always the same, which is why the acceleration is linear. When the dust has reached the same velocity as the shock, it is pushed into the slow flow of the pre-shock regime and the acceleration will abruptly come to an end. 

\begin{figure}[t!]
\center
\resizebox{\hsize}{!}{
\includegraphics[clip=true]{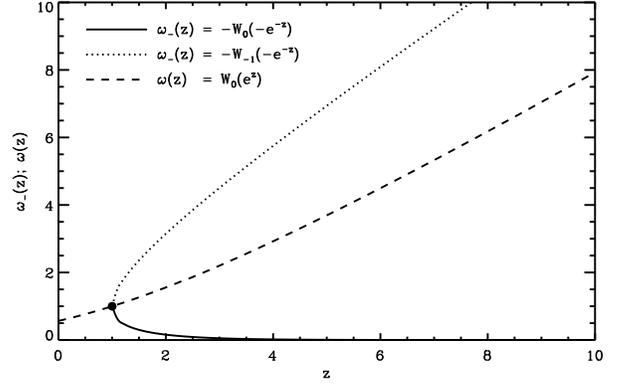}}
\caption{The original and the modified Wright omega functions for a real variable $z$. Note the double valued nature and that the function is undefined for $z\le 1$}.
\label{modomega}
\end{figure}

For the case $s < u_2 =$~constant one may change the dimensionless variables to
\begin{equation}
\varpi = {u_2-u_{\rm d}\over u_2-s} \ge 0,\quad \eta = {x-s\,t \over t_{\rm s}\,(u_2-s)},
\end{equation}
which yields a transformed equation of the form
\begin{equation}
\label{modwrighteq}
{d\varpi\over d\eta} = {\varpi\over \varpi-1},
\end{equation}
with the solution $\varpi(\eta) = \omega_-(\eta+C)$, where $\omega_-$ will here be referred to as the {\it modified Wright omega function}. This function, $ \omega_-$, is also related to the $W$ function and reflects the multi-valued nature of $W$ as it is constructed by the two real branches $W_0$ and $W_{-1}$. That is,
\begin{equation}
\omega_-(z) = \left\{
\begin{array}{lc}
 -W_0(-e^{-z}), & \omega_- \ge 1, \quad z \ge 1, \\[3mm]
 -W_{-1}(-e^{-z}), & \omega_- < 1, \quad z \ge 1, \\
 \end{array}
 \right.
\end{equation}
which also means that no solution exists for arguments less than one (see Fig. \ref{modomega}). In physical variables the solution becomes,
\begin{equation}
\label{physsol2}
u_{\rm d} = \omega_-\left[ {x-s\,t + x_0\over t_{\rm s} (u_2-s)}\right](s-u_2) + u_2,
\end{equation}
or,
\begin{equation}
\label{physsol3}
u_{\rm d} = \left\{
\begin{array}{lc} 
W_0(-e^\eta)(u_2-s) + u_2, & s \le u_1,\\[3mm]
W_{-1}(-e^\eta)(u_2-s)+ u_2, & u_1 < s < u_2,\\
 \end{array}
 \right.
\end{equation}
where
\begin{equation}
\eta = {x-s\,t + x_0\over t_{\rm s} (u_2-s)}.
\end{equation}
This solution does not exist beyond the point where $u_{\rm d} = s$, i.e., where $\omega_-(z)=1$ (note also that Eq. \ref{modwrighteq} has a singularity at $\varpi = 1$). The modified Wright omega function $\omega_- \to 1$ as $z\to 1$, but is not defined at $z \le1$. Thus, solutions exist only for $u_{\rm d} < s$. The physical interpretation of this is that the dust cannot be accelerated to velocities greater than the shock velocity $s$ because once the dust reaches the shock front, moving beyond it it would mean that the dust enters the slower pre-shock flow and experiences an opposite drag force (slowing down). The point where $u_{\rm d} = s$ is therefore also a kind of ``equilibrium point'', where the dust is repeatedly being ``kicked back'' as soon as the shock catches up and thus stays at $u_{\rm d} = s$.

\section{Examples}
\begin{figure}[t!]
\center
\resizebox{\hsize}{!}{
\includegraphics[clip=true]{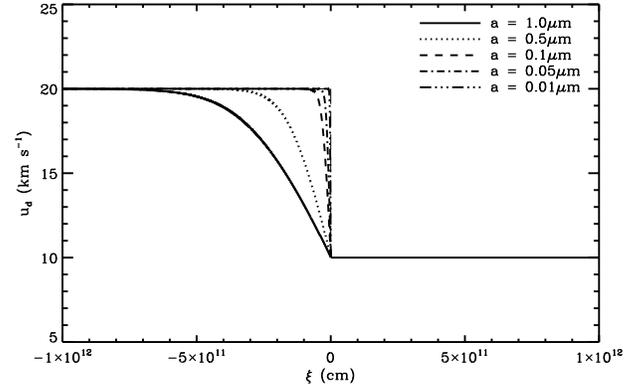}}
\caption{Solutions for $u_{\rm d}$ for a range of particle sizes and a shock velocity greater than the post-shock flow velocity. (The shock velocity is fixed to $s = 25$~km~s$^{-1}$.)
}
\label{ud_agr}
\end{figure}

Fig. \ref{ud_agr} show solutions for $u_{\rm d}$ for a range of particle sizes. The gas density (before the shock passage) and temperature is $\rho = 10^{-14}$~g~cm$^{-3}$ and $T= 100$~K ($v_{\rm th} \sim 1.0$~km~s$^{-1}$), respectively, which roughly corresponds to the conditions in the denser parts of the neutral ISM. The dust-to-gas ratio is assumed to be constant before the shock passage. Grains larger than $a\sim 0.1\,\mu$m appear to decouple from the flow for long enough to significantly influence the shock profile. 
The flow and shock velocities used in this example are $u_1 = u_{\rm d} = $~10~km~s$^{-1}$ before the shock, $u_2 = $~20~km~s$^{-1}$ after the shock and $s =$~25~km~s$^{-1}$. 

\begin{figure}[t!]
\center
\resizebox{\hsize}{!}{
\includegraphics[clip=true]{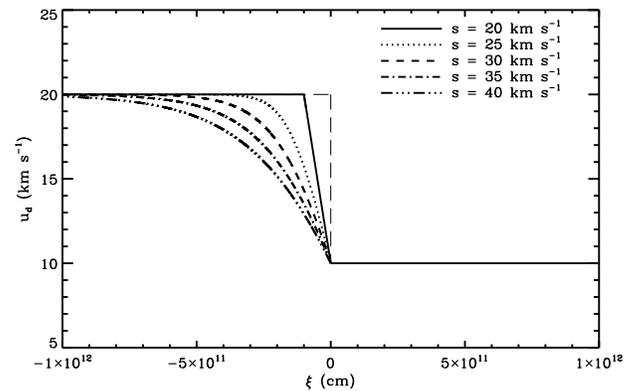}}
\caption{Solutions for $u_{\rm d}$ for shock velocities greater than the post-shock flow velocity and a fixed grain size $a = 0.1\,\mu$m.
}
\label{ud_s}
\end{figure}

The slower response of the dust to the propagating shockwave is due not only to the sizes of the dust grains, but also to the difference between the flow velocity after the shock $u_2$ and the shock velocity $s$. From Fig. \ref{ud_s} it is evident that the larger the difference, the slower the response of the dust to the shockwave. The reversal of this phenomenon may occur when a shockwave slows down. When $u_2=s$ the dust fluid has a linear response to the shock until $u_{\rm d} = u_2 = s$ is reached, which is due to a continuous ``kicking'' by the shock (see Section \ref{solutions}).

\begin{figure}[t!]
\center
\resizebox{\hsize}{!}{
\includegraphics[clip=true]{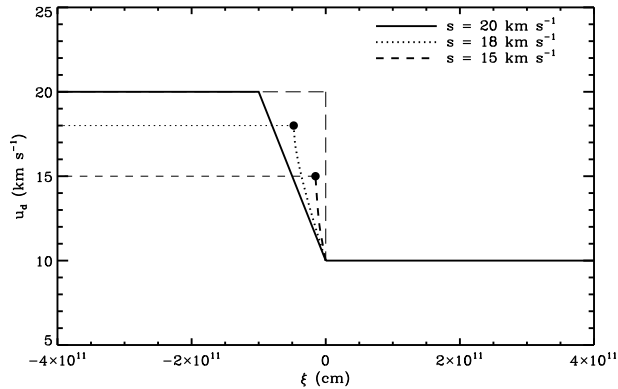}}
\caption{Solutions for $u_{\rm d}$ for shock velocities less or equal to the post-shock flow velocity and a fixed grain size $a = 0.1\,\mu$m. The filled black circles show the points where the solutions cease to exist.
}
\label{ud_s2}
\end{figure}

A step-like shock moving with a velocity which is in between the pre-shock flow velocity $u_1$ and the post-shock flow velocity $u_2$ is similar to the shock forming as a result of wind-wind interaction and leading to the formation of detached shells around C stars. In particular, this idealised step-shock is similar to the resultant profile in case C in \citet{Steffen00}. Fig. \ref{ud_s2} show a couple of examples where $u_1 < s < u_2$. When $u_{\rm d}$ reaches the shock velocity $s$, the dust would reenter the pre-shock region if $u_{\rm d}>s$ and thus be hit by the shock again. Hence, the dust cannot be accelerated to velocities above the shock velocity and will not connect to the post-shock flow. This means there is a permanent decoupling between gas and dust after the shock.

The dust- and wind-modelling results for carbon stars by \citet{Mattsson10} and \citet{Mattsson11} suggest the average radius of the grains formed in the winds of C stars (at solar metallicity) is $\langle a \rangle \sim 0.1\,\mu$m, or possibly even larger. Thus, the grains produced by these stars may be large enough for the assumption of velocity coupling between gas and dust in the circumstellar environment \citep[see, e.g.,][]{Mattsson07} to be unjustifiable. Detached shells, as results of wind-wind interactions, will likely show a broader profile \citep[compare, e.g.,][]{Olofsson90,Cox12} in dust than in gas emission and possibly also a slight spatial offset between the two, which remains consistent with observations \citep{Maercker10}.

\section{Test case for hydrodynamic gas/dust simulations?}
The idealised case considered above has some special features, which could make it useful as a test case for numerical simulations. For instance, the linear acceleration of the dust when the shock and post-shock flow velocities are equal ($s=u_2$) and the abrupt end of the acceleration when the dust reaches the shock velocity appears to be a challenge for any numerical scheme. So is also the case where the shock velocity is less than the post-shock flow velocity ($s<u_2$), since there exist no solution for the dust velocity for which $u_{\rm d} \ge s$. Moreover, numerical artefacts could also occur near $u_{\rm d} = s$. 

\section{Summary and conclusions}
An exact solution of a forced Burgers' equation representing the dynamics of a "dust fluid" in a one-dimensional flow has been presented. Adopting a constant stopping time the equation of motion for the dust fluid can be transformed into a simple ordinary differential equation, which is satisfied by the Wright omega function if the post-shock flow velocity $u_2$ is less than the shock velocity $s$. In case $s < u_2$ the solution is in terms of a related double valued function referred to as the modified Wright omega function. This function appears not to be previously described in the literature.

Because the response to passage of a shock wave is slower for dust than gas, the shock-velocity profile for the dust fluid is not the same as for the gas is. In case of typical interstellar conditions, grains larger than $0.1\,\mu$m will tend to decouple from the gas flow and the larger the difference between the post-shock gas velocity $u_2$ and the shock velocity $s$, the slower the response of the dust to the shockwave. 

If the shock velocity $s$ is larger than the post-shock gas velocity $u_2$, i.e., $s > u_2$, the dust will always catch up with the gas flow. How long this takes (or how much the shock-velocity profile is stretched/smoothened) depends on the sizes of grains as well as the velocity difference $s - u_2$. In the special case $u_2 = s$, the dust fluid is accelerated linearly, i.e., at a constant rate, until $u_{\rm d} = s$. In case $s < u_2$, the dust is accelerated at rate which increases until $u_{\rm d} = s$ and then the acceleration abruptly ends. $u_{\rm d} = s$ is thus an upper limit to the velocity of the dust fluid when $u_1 \le s\le u_2$, whereas the post-shock gas velocity $u_2$ is the upper limit when $s > u_2$. The dust-fluid flow will in the latter case always eventually couple to the gas flow, while in the former the two fluids will remain decoupled after the passage of the shock.

The simple analytical model discussed here displays distinct properties and may thus serve as a test case for numerical models of hydrodynamic gas-dust simulations where the dust fluid is coupled to the gas via, e.g., Epstein drag. A shockwave sent into an inviscid pressure-less gas will develop a step-like discontinuity which propagates like an idealised travelling wave (cf. the well--known travelling-wave solution to the Burgers' equation). A dust fluid coupled to this pressure-less gas (initially with velocity coupling between the two phases) will have a response to the passing shock which should be similar to analytical solutions derived in this work. 

\section*{Acknowledgments}
{The author wishes to thank Dhrubaditya Mitra for stimulating and enlightening discussions on the dynamics of particles in a fluid flow.}

\bibliographystyle{cs19proc}
\bibliography{agb.bib}

\end{document}